\documentclass[preprint,nofootinbib,superscriptaddress,amssymb]{revtex4}
\usepackage[pdftex]{graphicx}
\usepackage{amsmath}
\usepackage{amssymb}
\usepackage{tipa}
\usepackage{verbatim} 
\usepackage{enumitem} 
\usepackage{relsize} 
\usepackage{subfig}
\usepackage{dsfont}
\usepackage{hyperref}
\usepackage{mathtools} 

\begin{document}

\title{Detecting anomalies in CMB maps: a new method}
\author{Jayanth T. Neelakanta}
\affiliation{Department of Physics, Syracuse University, Syracuse, NY 13244-1130, USA}

\begin{abstract}
\noindent Ever since WMAP announced its first results, different analyses have shown that there is weak evidence for several large-scale anomalies in the CMB data. While the evidence for each anomaly appears to be weak, the fact that there are multiple seemingly unrelated anomalies makes it difficult to account for them via a single statistical fluke. So, one is led to considering a combination of these anomalies. But, if we ``hand-pick" the anomalies (test statistics) to consider, we are making an \textit{a posteriori} choice. In this article, we propose two statistics that do not suffer from this problem. The statistics are linear and quadratic combinations of the $a_{\ell m}$'s with random co-efficients, and they test the null hypothesis that the $a_{\ell m}$'s are independent, normally-distributed, zero-mean random variables with an $m$-independent variance. The motivation for such statistics is generality; equivalently, it is a non \textit{a posteriori} choice. But, a very useful by-product of considering such statistics is this: Because most physical models that lead to large-scale anomalies result in coupling multiple $\ell$ and $m$ modes, the ``coherence" of this coupling should get enhanced if a combination of different modes is considered. Using fiducial data, we demonstrate that the method works and discuss how it can be used with actual CMB data to make quite general statements about how incompatible the data are with the null hypothesis.
\end{abstract}

\maketitle

\section{Introduction}

A dramatic increase in the amount of observed data has, over the last couple of decades, led to a much better understanding of the Universe we inhabit. In fact, the cosmology community is so confident about the standard paradigm that the paradigm is referred to as the Standard (or Concordance) Model, after the Standard Model of Particle Physics. Seven or eight parameters, along with general relativity and elementary quantum mechanics, are sufficient to explain a host of observations on the largest scales, once initial conditions are set deep in the radiation era. Standard field quantization techniques applied to cosmic inflation have been remarkably successful in explaining these initial conditions even. The cosmology being studied today is called Precision Cosmology because parameters have been determined to percent-level precision \cite{Hinshaw:2012aka,Ade:2013zuv}.

But, as is well-known, there is a difference between precision and accuracy. Questions abound over some of the postulates of the Concordance Model. Because we have access to only one universe, the usual method of testing postulates by repeating experiments cannot be carried out. As inflation postulates that the primordial seeds of the universe's structure themselves arise out of a stochastic process, this inability to repeat experiments is an even bigger handicap.

The cosmic microwave background (CMB) has turned out to be the cosmologist's most useful aid in understanding what has happened in the universe from just a few minutes after the Big Bang, all the way up to the present. Since most CMB photons have travelled to us without any scattering, they represent a very faithful picture of the universe when it was about 400,000 years old. Moreover, at the scales relevant to us today, the density perturbations were small enough that linear perturbation theory is an excellent approximation. This implies that the statistical properties of the primordial fluctuations were preserved all the way to the surface of last scattering, and thence to us today.

In vanilla models of inflation, the Fourier modes of the primordial fluctuations have the same dynamics as harmonic oscillators in their ground state, and are thus distributed as Gaussians \footnote{We ignore non-Gaussianities of the kind calculated by Maldacena \cite{Maldacena:2002vr} as they are highly suppressed.}. Moreover, statistical homogeneity and isotropy imply that the variance of this Gaussian distribution doesn't depend on the direction of the wavenumber of the Fourier mode, and that the variance is the same for the real and the imaginary part of the Fourier modes \cite{Liddle:2000cg}. In 2013, Planck announced that the CMB data put very strong constraints on the amount of non-Gaussianity in the primordial power spectrum \cite{planck_ng}. In effect, this meant that several exotic inflationary models got ruled out with very high probability. So, the accepted wisdom is that the Fourier modes of the initial density perturbations are independent and distributed as Gaussians.

The only challenge to this postulate of independent, normally distributed perturbations probably has to do with the so-called CMB anomalies. The CMB anisotropies across the sky are usually expressed in terms of $a_{\ell m}$'s, the co-efficients corresponding to the spherical harmonics $Y_{\ell m}$'s. Expressing Fourier modes in terms of the spherical harmonics, and using the results from the previous paragraph, we are led to conclude that most viable inflationary models predict that the $a_{\ell m}$'s are normally distributed with zero mean and with a variance $C_{\ell}$ (hence independent of $m$).

When WMAP announced \cite{Hinshaw:2003ex} its first set of results, the authors in \cite{Tegmark:2003ve} and \cite{oliviera} analysed the $a_{\ell m}$'s to test this hypothesis. They employed a variety of tests and found weak evidence for correlation amongst the $a_{\ell m}$'s that corresponded to the largest scales (low-$\ell$'s). The anomalies reported dealt with the alignment of different multipoles and how planar a few of these multipoles were. Several authors \cite{copi,Schwarz:2004gk,Bielewicz:2005zu}  performed similar analyses and again found weak evidence. A different kind of anomaly, having to do with a low value for the variance in the CMB sky, was observed by \cite{Monteserin:2007fv} for the WMAP data. The authors in \cite{Hansen:2004vq,Park:2003qd,Eriksen:2003db} considered the isotropy of the angular power spectrum and concluded that it appears to be anisotropic. A few more anomalies were reported in \cite{Vielva:2003et,Land:2005jq,Hansen:2008ym}, amongst other works.

Apart from the weak evidence, two arguments were proferred questioning the ``real" nature of these anomalies: one, that they arose from the systematics that WMAP employed; two, that these anomalies were checked for \textit{a posteriori}. So, now that Planck has confirmed that most of the anomalies are present in their data too \cite{planck}, one may reasonably argue that the anomalies are a \textit{bona fide} feature of the CMB. The question remains as to whether this feature is physically relevant or not. As Planck also concluded that there is only weak evidence for these anomalies, this question has not been settled convincingly. Many authors contend \cite{wmap_stats}, with good reason, that given a large enough dataset, one can always find any feature that one desires. Compounding the problem of the large dataset is the fact that the anomalies have been observed for low-$\ell$'s\textemdash it is here that the effect of cosmic variance is most pronounced. This makes statistical inferences about the anomalies even more dubious. Also, there is the perennial question of foreground contamination\textemdash without a reliable model for galactic dust, it isn't clear how accurate the determination of the $a_{\ell m}$'s is. (Though, with the availability of multiple probes and multiple frequency channels, this is less of an issue \cite{Bobin:2014mja} than it used to be.)

But, the fact remains that there are many anomalies with weak evidence. Some of them are so apparently different from the rest that, at the outset, it is seems hard to believe that they all arose from a common statistical fluke. And, the anomalies seem to be present ``coherently" across different $\ell$'s too, seemingly making it harder to believe that it is the consequence of a fluke.

This paper tries to address the second of the arguments put forth against the existence of the anomalies\textemdash that the tests are all \textit{a posteriori}. We propose two statistics that test the null hypothesis that the $a_{\ell m}$'s are independent, normally distributed zero-mean variables. As we shall show, these statistics are such that one cannot reasonably be accused of performing the analysis after ``seeing" the anomalies in the data. The point is to perform as general an analysis of the data as possible, without worrying about whether a test statistic is physically-motivated or high-confidence-interval motivated. We shall achieve this by not arbitrarily choosing the $\ell$'s and the $m$'s to analyse; instead, we consider their linear and quadratic combinations. For one, this makes the analysis more general; but, crucially, if the anomalies are physical, it is very unlikely that they arose because of a coupling between just two or three $a_{\ell m}$'s. This anomalous nature must be present for a range of modes and thus considering combinations of the modes should lead to an enhancement of the signal. Also, previous analyses of CMB anomalies have involved several Monte Carlo simulations to produce a reference set of Gaussian sky maps. And, one gets several $p$-values as different statistics are considered. In our case, once a maximum $\ell$ value is chosen, one gets a single $p$-value for each of the two statistics considered.

Of course, the real interest lies in applying this test to actual data. But, in this article, we restrict ourselves to first demonstrating that this test actually works. This is partly due to the fact that this is a novel method and we would like to introduce it and test its usefulness; it is also due to the fact that different missions and different foreground subtractions have led to a plethora of maps. Hopefully, in a subsequent article, we will apply the proposed test in a manner consistent with the multiplicity of the available maps. Also, a Planck release of the polarization data is expected in the near future. It would be interesting to test the statistical properties of this too, along with that of the scalar modes.

\section{Y\textemdash A Linear Statistic}

\subsection{Motivation}

In broad terms, the way a hypothesis is statistically tested is this: Assume that a given dataset is described by a known probability distribution $P_1$; formulate a statistic that is a function of the corresponding random variables; determine the expected distribution $P_2$ of this statistic, assuming the fiducial distribution $P_1$; see how compatible the actual (realized) value of the statistic using the given dataset is, with the distribution $P_2$. If the compatibility is very low, then one concludes that the data are inconsistent with the hypothesis.\footnote{This is more of a goodness-of-fit test than a hypothesis test because we are not specifying an alternative hypothesis. But, the former can be thought of as a special case of the latter, where the alternative hypothesis is that the data are \textit{not} described by the null hypothesis.} It is clear, however, that the conclusion strongly depends on the statistic chosen. Ideally, one would like to do the analysis for several different statistics.

Let us look at linear test statistics; that is, if an $n$-component vector $\vec{X}$ describes $n$ variables of a dataset, then consider \(S=\vec{a}\cdot\vec{X}\), where each choice of the constant co-efficients $\vec{a}$ would correspond to one statistic. If one wants to do a blind analysis of the data, one is tempted to consider several different choices of $\vec{a}$\textemdash for instance, by making $\vec{a}$ itself a random vector. If one knows the underlying distribution of $\vec{a}$, and the null hypothesis for the distribution of $\vec{X}$, then one may hope to determine the distribution of $S$. In general, this distribution would be quite complicated. In the next section, we show that for a specific choice of the distribution of $\vec{a}$, and a specific null hypothesis, the distribution of $S$ becomes very simple.

\subsection{The $Y$ Statistic}\label{subsec:y_ensemble}

Let $\vec{a}$ be an $N$-component random variable vector, with each component being described by a zero-mean normal distribution, $\mathcal{N}\left(0,\alpha_i^2\right)$. Let $\vec{X}$ be another $N$-component vector with each of its components being described by $\mathcal{N}\left(0,\beta_i^2\right)$. Further, assume that \(\alpha_i^2 = 1/\beta_i^2\). That is, the combined probability distribution function is
\begin{align}\label{eqn:pdf_a_X}
P(\vec{a}, \vec{X}) = \frac{1}{(2\pi)^N} \left(\det \boldsymbol{\Sigma_a}\,  \det \boldsymbol{\Sigma_X}\right)^{-1/2} \, \, \exp \left(-\frac{1}{2} \vec{a}^{\,\,T} \,\, \boldsymbol{\Sigma_a^{-1}} \, \, \vec{a}\right) \exp \left(-\frac{1}{2} \vec{X}^{\,\,T} \,\, \boldsymbol{\Sigma_X^{-1}} \, \, \vec{X}\right),
\end{align}
where $\boldsymbol{\Sigma_a}$ and $\boldsymbol{\Sigma_X}$ are diagonal matrices with \(\left(\boldsymbol{\Sigma_a}\right)_{ij} = \left(\boldsymbol{\Sigma_X}\right)^{-1}_{ij}  =  \alpha_i^2 \delta_{ij}\).

Consider a random variable arising out of these two random variables,
\begin{align}\label{eqn:defn_Y}
Y = \vec{a} \cdot \vec{X} = a_i X^i 
\end{align}
In the above, Einstein's summation convention is implied. Though both $\vec{a}$ and $\vec{X}$ are random variables, we shall eventually consider the case where there is only one realization of $\vec{X}$. That is, the two random variables must not be considered to be on the same footing. We shall first treat $\vec{X}$ as a constant vector, carry out all operations with respect to $\vec{a}$ and finally promote $\vec{X}$ to a random vector and carry out operations with respect to it. This shall become more clear when we apply it to the case of Cosmology.

For a given realization of $\vec{X}$, $Y$ is a linear combination of the independent normal variables $\vec{a}$. Hence, $Y$ is normally distributed too: 
\begin{align} \label{eqn:sigma_def}
Y \sim \mathcal{N}\left(0, \alpha_1^2 X_1^2 + \dots + \alpha_N^2 X_N^2\right) \coloneqq  \mathcal{N}\left(0, \sigma^2 \right)
\end{align}
This is for a given realization of the $X^i$'s. But, the $X^i$'s themselves are random variables with an underlying distribution. Thus, we may ask how $\sigma^2$ is distributed. Because \(\alpha_i^2 = 1/\beta_i^2\), that is, the reciprocal of the variance of $X^i$, $\sigma^2$ is the sum of squares of $N$ normally distributed random variables with zero mean and unit variance. Hence, $\sigma^2$ follows a Chi-squared distribution with $N$ degrees of freedom, \( \sigma^2 \sim \chi^2(N)\). To calculate the probability distribution of $Y$, that is, $P(Y=y)$, we need to marginalize over this $\chi^2(N)$ because the variance is now a random variable:
\begin{align}
P(Y=y) &= \int_{0}^{\infty} \mathrm{d}\sigma^2  \, \, P(y|\sigma^2)  \, P(\sigma^2) \notag \\
&= \int_{0}^{\infty} \mathrm{d}\sigma^2 \frac{1}{\sqrt{2 \pi \sigma^2}} \exp \left[\frac{-y^2}{2 \sigma^2} \right] \chi^2(N,\sigma^2) \notag && \text{[From (\ref{eqn:sigma_def})]}\\
&= \sqrt{\frac{1}{\pi}\left(\frac{|y|}{2}\right)^{N-1}} \quad \frac{\mathrm{K}\left(\frac{N-1}{2}, |y|\right)}{\Gamma\left(N/2\right)}\label{eqn:dist_y_final},
\end{align}
where $\mathrm{K}$ is the modified Bessel function of the second kind.\footnote{This distribution can be considered to be the generalization of the well-known distribution of the random variable that is the product of two standard Gaussian variables. The latter corresponds to the \(N=1\) case of the former.}

As the dependence of $Y$ is only on $N$, one may wonder where the distributions of $\vec{a}$ and $\vec{X}$ enter. It is only because of the choice of the variances of the distribution, \(\alpha_i^2 = 1/\beta_i^2\), that the dependence on the details of the distribution ``cancels out". So, as promised earlier, we have shown that a specific choice of the distribution for the co-efficients ($\vec{a}$ in this case) results in a very simple form for the distribution of the statistic.

Usually, the word (test) statistic is reserved for a function of the data. In particular, for each set of data, such a statistic returns a single number. In our case, by construction, the $Y$ statistic is not a single number because a given $\vec{X}$ is multiplied by several $\vec{a}$. We shall call such statistics \textit{vector statistics}.

\subsection{Realizations}

In cosmology, we have only one realization of the Universe. For our purposes, this translates into one realization of the $a_{\ell m}$'s, which we take to be the \textit{real} multipole co-efficients corresponding to the real spherical harmonics $Y_{\ell m}$'s (see, for instance, Appendix A of \cite{cristian_mean}). The index $m$ then ranges from \([-\ell,\ell]\). But, the Concordance Model of Cosmology predicts that each $a_{\ell m}$ is a random variable, arising from a Gaussian distribution $\mathcal{N}(0,C_{\ell})$. This can be thought of as the null hypothesis $H_0$.

Thus, the $a_{\ell m}$'s are like our $\vec{X}$ and we shall refer to them as $\vec{X}$, in order to keep matters general. One way of testing $H_0$ is by considering different test statistics of $\vec{X}$ and seeing if their realized values are compatible with that predicted by the null hypothesis. One trouble with this method is that $\vec{X}$ doesn't have a basis-independent definition\textemdash its meaning depends on the coordinate system employed in the sky. Further, the $p$-values that one derives depend fundamentally on the test statistic chosen. So, just because one such $p$-value is compatible with the null hypothesis doesn't mean that the data are.

In our case, note that the $Y$ statistic is independent of the co-ordinate system chosen in the sky. To see this, consider an active rotation of the co-ordinate system. It can be shown that the transformed $\vec{X}$, say $\vec{X}'$, is related to $\vec{X}$ by a real orthogonal matrix\footnote{The transformation matrix is given by  $\mathcal{C}^*  \mathcal{D} \,  \mathcal{C}^{\, T}$ \cite{Blanco199719}, where $\mathcal{C}$ is a matrix that relates the complex spherical harmonics to the real ones, and $\mathcal{D}$ is the Wigner D-matrix \cite{wigner} that describes how complex spherical harmonics transform under rotations. Both matrices are unitary and $^*$ denotes complex conjugation.}, say  $\mathcal{R}$; that is, \(\vec{X}' = \mathcal{R} \cdot \vec{X}\). The $Y$ statistic arising out of $\vec{X'}$, say \(Y' = a_i (X')^i = a_i \mathcal{R}^i_{\, \,k} X^k =  \mathcal{R}^{\, \, \, \,i}_k a_i X^k \coloneqq (a')_i X^i\). Using (\ref{eqn:pdf_a_X}) and \(\mathcal{R}^{\, T} \mathcal{R} = \mathds{1}\), it is clear that the PDFs for $a_i$ and $(a')_i$ are the same and hence the $Y$ statistic is co-ordinate system independent.

Now that we have discussed the test statistic and its properties, let us detail our motivation for considering this statistic and what we intend to do with it. One might wonder why a linear combination of the components of $\vec{X}$ is being considered. This has to do with the kind of anomalies that are usually discussed. It is very natural to assume that these anomalies are the result of some correlation between the different components of $\vec{X}$. Indeed, many models that attempt to explain these anomalies posit precisely such a correlation (see \cite{Copi:2010na} and references therein for a review of the anomalies and some proposed explanations for their origins). The only way to test these correlations is by considering functions that ``mix" the different components. A linear superposition is just the simplest of these functions. We shall consider second-order statistics in due course.

We now consider a more operational definition of $\vec{X}$. We specialize to the case where $\vec{x}$, the realization of $\vec{X}$, is the set of $a_{\ell m}$'s. That is \(x_1=a_{2,-2}, \, \, x_2=a_{2,-1}, \dots , x_5=a_{2,2}; \, \, x_6=a_{3,-3}, \dots; x_N=a_{\ell_{\mathrm{max}},-\ell_{\mathrm{max}}}\).\footnote{As is usual in CMB analyses, we ignore the monopole and the dipole ($\ell=0$ and $\ell=1$).} Here, $\ell_{\mathrm{max}}$ is the largest $\ell$ value that we go up to:
\begin{align} \label{eqn:lmax}
\ell_{\mathrm{max}}^2 + 2 \ell_{\mathrm{max}} - (3+N) = 0
\end{align}

The strategy is the following: Under the null hypothesis $H_0$, we have the distribution for $Y$, given in (\ref{eqn:dist_y_final}). From CMB experiments such as WMAP and Planck, we have the realized values of $\vec{X}$ in the actual sky. We use these realized values of $\vec{X}$, $\vec{x}_{sky}$, and determine the distribution of $Y$, $P(y_{sky})$. We can compare this distribution with (\ref{eqn:dist_y_final}) and can then infer the compatibility of CMB data with $H_0$.

\section{Hypothesis Testing}\label{sec:hyp_y}

Usually, hypothesis testing involves calculating the probability of the realized value of a statistic, given the distribution of the statistic under the assumption of the null hypothesis. This procedure cannot be directly implemented in our approach because, by construction, our test statistic $Y$ doesn't yield a single number for a given dataset - it is a vector statistic. So, whereas in the usual case we only have to compare one realized value of the test statistic with the expected value, in our case, by its very nature, we must compare the realized distribution $P(y_{sky})$ with that in (\ref{eqn:dist_y_final}). 

Now, there is no unique way of comparing two arbitrary distributions. As we are basically looking for a measure of goodness-of-fit, we could consider a chi-squared test. But, chi-squared tests are more useful in circumstances where one is estimating the parameters in a given model. In that case, minimising chi-squared leads to the best-fit parameters. That is not what we are doing here. We are actually comparing data with a fiducial distribution function. Moreover, using the chi-squared test involves binning the data, and some information is lost in this process. It would be more desirable to work with tests that use the data themselves, not bins of data.

Different such tests have been proposed in the literature, and we shall adopt the Anderson-Darling (A-D) test \cite{A-D}, which we shall describe shortly. The reason for the choice is that studies \cite{stephens} have shown that, for a variety of distributions, this test is more powerful than others such as the more commonly used Kolmogorov-Smirnov (K-S) test. A possible drawback of using the A-D test instead of, say, the K-S test is that the critical values depend on the distribution corresponding to the null hypothesis, but, because we know the form of this distribution (\ref{eqn:dist_y_final}), the critical values can be calculated. Moreover, this dependence on the distribution is reflective of the fact that the A-D test is much more sensitive to the underlying distribution than the K-S test, and hence more powerful.

\subsection{Anderson-Darling Test} \label{sec:A-D}

Let $V$ be a random variable and let the null hypothesis be that the (continuous) probability distribution $F(V)$ describes this variable. Further, let the $m$-component vector $v_i$ represent $m$ samples of $V$, sorted in increasing order. Define $\Phi(w)$ to be the cumulative distribution function,
\[
\Phi(w) = \int_{-\infty}^{w} dv \, F(v)
\]
Also, define 
\begin{align} \label{eqn:A-D S}
S = \frac{1}{m}\sum_{i=1}^{m} (2i-1) \Big(\log[\Phi(v_i)] + \log[1-\Phi(v_{m+1-i})]\Big)
\end{align}
The A-D statistic is then given by 
\begin{align} \label{eqn:A-D stat}
A^2 = -m - S
\end{align}

For well-known distributions, such as the normal distribution, critical values of $A^2$ statistic have been calculated in the literature. Associated with each critical value is a $p$-value, with which the null hypothesis can be rejected at the corresponding significance. For example, a value of $A^2$ more than 3.857 would mean rejecting the null hypothesis that the data are described by a normal distribution with a given mean and variance at the 1\% level.\footnote{This is in the limit of infinite data, and for data that have been standardised (subtract the mean from the data, and divide by the standard deviation), though, for the case of the normal distribution, modifications for finite $m$ exist.}

As our distribution (\ref{eqn:dist_y_final}) is not one of the common distributions (the earliest reference to it that we could find is in \cite{pearson}), published critical values for the A-D test do not exist. But, for a given $N$, we can determine them simply by generating a large number of realizations drawn from (\ref{eqn:dist_y_final}), calculating the corresponding value of $A^2$, and repeating this procedure a sufficient number of times. This would give us the distribution of $A^2$ for (\ref{eqn:dist_y_final}), from which the critical values can be calculated. Call this distribution $\Psi_Y(A^2, N)$.

A peculiar feature arises out of the fact that we only have access to one realization of the $a_{\ell m}$'s. For typical PDFs, the distribution of $A^2$ in (\ref{eqn:A-D stat}) asymptotes fairly quickly to a fixed distribution as the number of realizations ($m$ in the equation) increases. But, recall that we have only one realization of $\vec{X}$. So, even if we increase the number of $Y$ statistics generated (thereby increasing the corresponding $m$), this is not equivalent to an ergodic sampling of the distribution. In particular, if we choose, say, \(m=10^5\), then, it \textit{does} matter whether we generate $m$ realizations of $Y$ by choosing $10^5$ realizations of $\vec{a}$ and $1$ realization of $\vec{X}$, or by choosing $10^3$ realizations of $\vec{a}$ and $10^2$ realization of $\vec{X}$. Thus, it turns out that in our case the distribution of $A^2$ for a given $N$, what we called $\Psi_Y(A^2, N)$, depends on $m$. We shall denote this distribution by $\Psi_Y(A^2, N, m)$. Implicit in this notation is the fact that we are choosing only one realization of $\vec{X}$.

\begin{figure}[t]
    \centering
    \subfloat{{\includegraphics[width=0.6\linewidth]{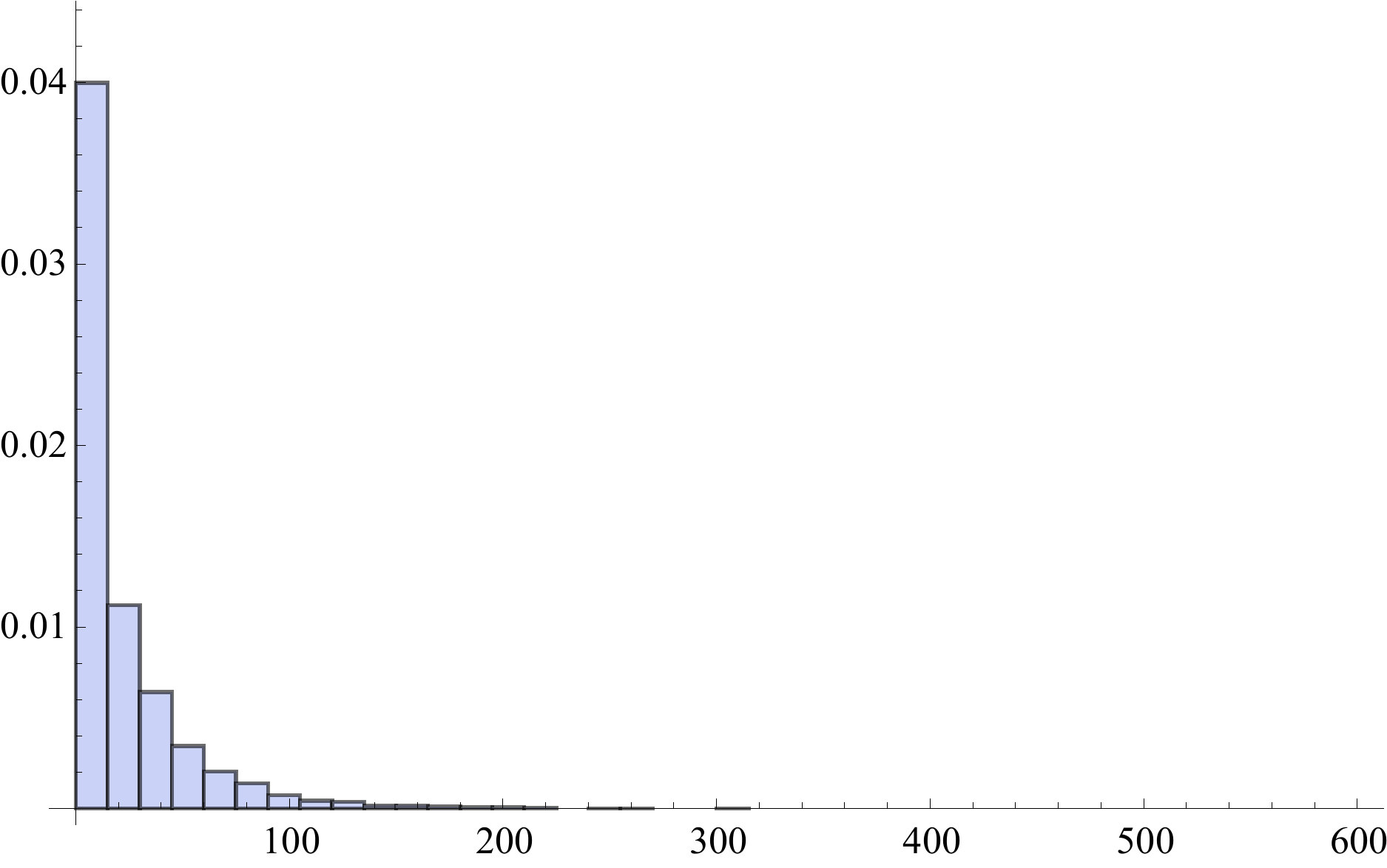} }}
    \quad
    \caption{PDF for $\Psi_Y(A^2, N, m)$ for \(m=10^5, \, N=672\).}
    \label{fig:pdf_psi_Y}
\end{figure}

For a given $N$ and $m$, we can determine $\Psi_Y(A^2, N, m)$ numerically by simply following the procedure outlined in (\ref{eqn:defn_Y}): We choose both $a$ and $X$ to be $N$-dimensional normal vectors with zero mean and unit variance. We pick one realization of $\vec{X}$ and $m$ realizations of $\vec{a}$, calculate the corresponding $Y$ and one corresponding realization of $\Psi_Y(A^2, N, m)$. We repeat this procedure several times until we have mapped out the distribution $\Psi_Y(A^2, N, m)$ reasonably well. This distribution for \(N=672\) and \(m=10^5\) is shown in Figure \ref{fig:pdf_psi_Y}.

Once we have determined $\Psi_Y(A^2, N, m)$, putting limits on how anomalous the data are, in terms of our formalism, is relatively straightforward. We have discussed in the previous section how we can generate a given number (say, $m$) of the $Y$ statistic. On sorting this data vector in increasing order, we can proceed to calculate the realized value of $A^2$, with the distribution in (\ref{eqn:dist_y_final}) corresponding to $F(V)$ above. Call this $a_{\text{sky}}^{2}$. This value can be compared with $\Psi_Y(A^2, N, m)$ and a $p$-value can be calculated.

\section{Z\textemdash A Quadratic Statistic}

\subsection{Ensemble}

In the previous section, we considered a linear combination of the different $a_{\ell m}$'s. We mentioned that if the anomalies are real, then models that could produce these anomalies without correlations amongst the different $a_{\ell m}$'s would likely be very contrived. So, to probe these correlations better, it is natural to consider test statistics that are second order in the $a_{\ell m}$'s. We shall do that here and revert to using $\vec{X}$ to denote the ordered set of $a_{\ell m}$'s.

Consider the following test statistic:
\begin{align}\label{eqn:z_def}
Z = B_{12} X_1 X_2 + B_{34} X_3 X_4 + \dots + B_{N-1  N} X_{N-1} X_N 
\end{align}
For now, assume that $N$ is even, so that this definition always makes sense ($N$ is even for an odd $\ell_{\mathrm{max}}$). We shall comment on dealing with an odd $N$ later.

$B_{ij}$ is a random variable distributed as $\mathcal{N}\left(0,\sigma_{Bij}^2\right)$, where \(\sigma_{Bij}^2=\frac{1}{X_i^2 \beta_j^2}\). Recall that $\beta_j^2$ is the variance of the normally distributed $X_j$. Note that we are using $X_i$ itself as a parameter describing a distribution. (Compare this with the distribution of $\vec{a}$, which depended on the variance of $\vec{X}$, and not on $\vec{X}$ itself.) This is not an issue because $\vec{X}$ is still being treated as a fixed vector. The reason for this choice of $\sigma_{Bij}^2$ will become clear momentarily, but, it must be borne in mind that it gets determined \textit{after} a choice of $\vec{X}$ is made.

With this, $Z$ is basically a sum of $N/2$ Gaussian random variables $B_{ij}$, with constant coefficients $X_i X_j$. Thus, we have that $Z \sim \mathcal{N}\left(0,\sigma_{Z}^2\right)$, where
\begin{align*}
\sigma_{Z}^2 &= X_1^2 X_2^2 \sigma_{B12}^2 + \dots + X_{N-1}^2 X_{N}^2 \sigma_{B, N-1, N}^2\\
&=\frac{X_2^2}{\beta_2^2} + \dots + \frac{X_N^2}{\beta_N^2}\\
&\sim \chi^2(N/2).
\end{align*}

Here, similar to the analysis in Section \ref{subsec:y_ensemble}, we have used the fact that $\sigma_{Z}^2$ is the sum of the squares of $N/2$ normally distributed, zero-mean random variables with unit variance. Though this has been said several times already, because of the novel nature of this treatment, it must be stressed that, up to now, $\vec{X}$ has been treated as a fixed vector. 

Similar to what we did for the test statistic $Y$, we now perform an ensemble average of $Z$ with respect to the distribution of $\vec{X}$. Repeating the calculation that led to (\ref{eqn:dist_y_final}), with half the number of terms, we have that the distribution of $Z$ is
\begin{align}\label{eqn:dist_z}
P(Z=z) = \sqrt{\frac{1}{\pi}\left(\frac{|z|}{2}\right)^{(N/2-1)}} \quad \frac{\mathrm{K}\left(\frac{N-2}{4}, |z|\right)}{\Gamma\left(N/4\right)}
\end{align}

Again, because of the choice of the distribution of the $B_{ij}$ variables, the distribution of $Z$ is solely a function of $N$. This is quite a useful feature for the following reason: Consider four random variables $R_1, R_2, R_3, R_4$. Let only $R_1$ and $R_3$ be correlated, and $R_2$ and $R_4$ be correlated:
\begin{align}\label{eqn:corr_Rs}
\langle R_1 R_3 \rangle = \langle R_2 R_4 \rangle = \epsilon, \, \text{where } \epsilon \ll 1
\end{align}
Now, say you are testing the null hypothesis that all four variables are mutually independent. You come up with two test statistics, $T_1 \coloneqq R_1 R_2 + R_3 R_4$ and $T_2 \coloneqq R_1 R_3 + R_2 R_4$. From (\ref{eqn:corr_Rs}), it is clear that $\langle T_1 \rangle$ is indistinguishable from that predicted by the null hypothesis, whereas $\langle T_2 \rangle$ gives a different prediction from the null hypothesis. Of course, the distribution of both $T_1$ and $T_2$ will be different from that predicted by the null hypothesis, but, at least for non-pathological distributions, $T_1$ is an $\mathcal{O}(\epsilon)$ worse discriminator for testing the null hypothesis.

If the CMB anomalies are due to correlations amongst the different $a_{\ell m}$'s, from the form of (\ref{eqn:z_def}), one may na\"{i}vely worry that just like with the $R_i$'s, the order in which the $a_{\ell m}$'s appear in the equation may matter. That is, instead of the order in (\ref{eqn:z_def}), one could alternatively consider 
\[
Z' = B_{13} X_1 X_3 + B_{24} X_2 X_4 + \dots + B_{N-2,N} X_{N-2} X_N 
\]
This is a different statistic from $Z$. In this manner, there are $(N-1)!!$ alternatives\footnote{Consider the sequence $\{1..N\}$. Each index has to occur once. So, there is no freedom in choosing the $i$ in (\ref{eqn:z_def}). For the $j$ corresponding to the first term, there are $(N-1)$ possibilities. Again, the $i$ for the second term is effectively fixed, as it must appear in the sum. For the $j$ corresponding to this term, there are $(N-3)$ possibilities, and so on.} to $Z$. In principle, each of these combinations will have a different distribution for $Z$. But, because of our choice of the distribution of $B_{ij}$, we have that the distribution of $Z$ depends only on $N$. With this motivation, let us define $\mathrm{Perm}(Z)$ as a permutation of the indices in $Z$ that ensures that each index appears once and only once. Now, define $\widetilde{Z}$ as the set of all $\mathrm{Perm}(Z)$. It is obvious that $\widetilde{Z}$ is distributed as (\ref{eqn:dist_z}). It is $\widetilde{Z}$ that is the statistic that we shall consider for the rest of this article, though, by abuse of notation, we shall refer to it as $Z$. In this way, the choice of the distribution function for $B_{ij}$ helps us overcome the difficulty of having to consider $(N-1)!!$ different distributions, while ensuring that there is no loss of generality in the sequence of indices chosen.

Finally, we had earlier stated that we would talk about the case with an odd $N$, which arises if we have an even $\ell_{\mathrm{max}}$. In that case, we can just consider pairs of the first $(N-1)$ of the indices of Perm(\(\{1,\dots,N\}\)), which occurs in $\widetilde{Z}$ anyway. This would mean that we are losing out on one mode during every permutation, but, the procedure ensures that there isn't any arbitrariness in the choice of that mode.

\subsection{Hypothesis Testing}

\begin{figure}[t]
    \centering
    \subfloat{{\includegraphics[width=0.6\linewidth]{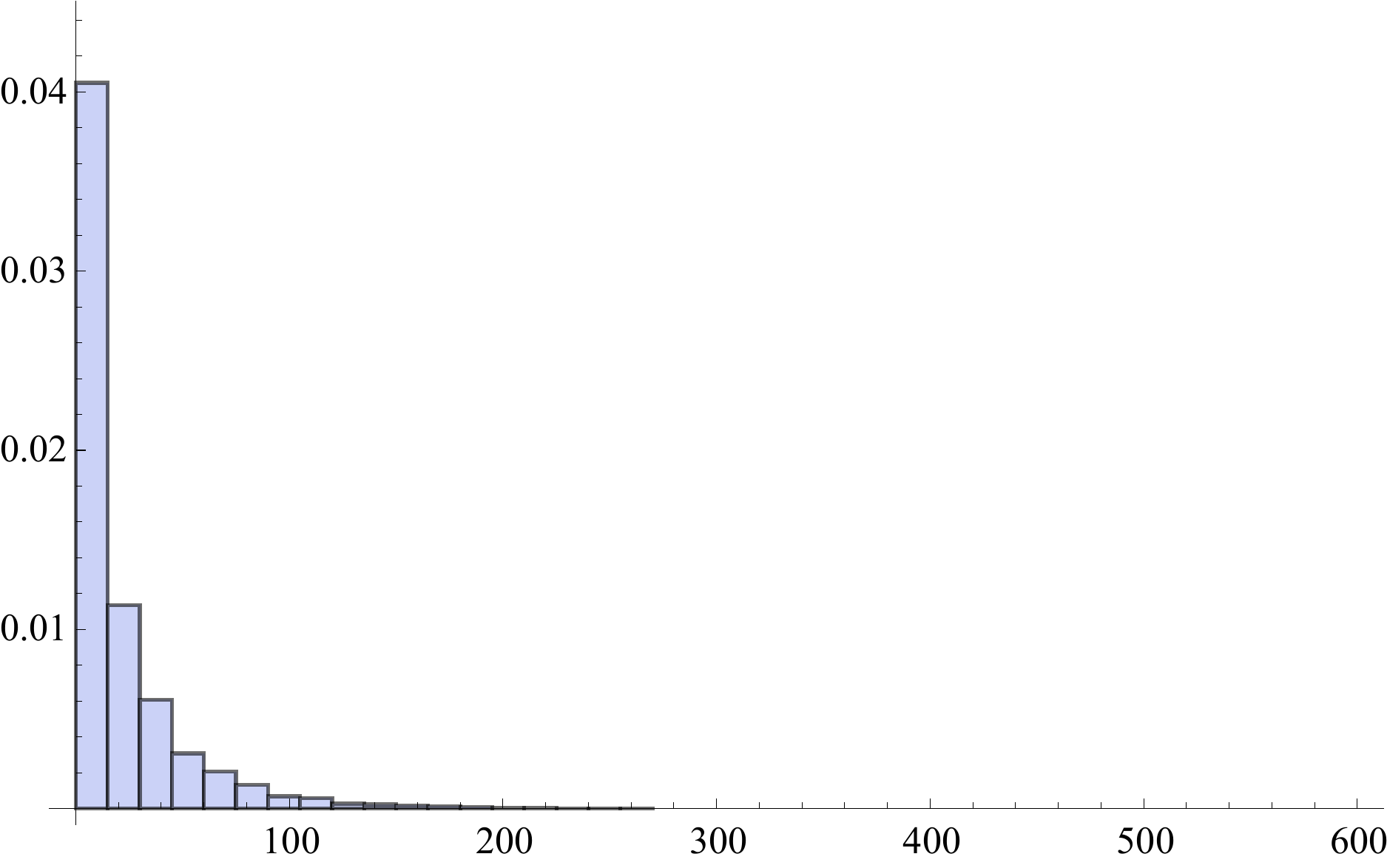} }}
    \quad
    \caption{PDF for $\Psi_Z(A^2, N, m)$ for \(m=10^5, \, N=672\).}
    \label{fig:pdf_psi_Z}
\end{figure}

The procedure of testing the null hypothesis $H_0$ is identical to the one we employed for the linear test statistic $Y$. The expected distribution under $H_0$ is given by (\ref{eqn:dist_z}) and we can use actual data to determine the realized distribution. We can then calculate the statistical significance of a departure from $H_0$ by using the procedure outlined in the previous section. Let us denote the probability distribution function for the Anderson-Darling statistic for the $Z$ statistic  as $\Psi_Z(A^2, N, m)$. We can repeat the procedure outlined in \ref{sec:A-D} to determine this PDF\textemdash the only change will be that, instead of generating distributions of the $Y$ statistic, we will generate distributions of the $Z$ statistic, and, instead of using (\ref{eqn:dist_y_final}), we shall use (\ref{eqn:dist_z}). For a particular choice of $N$ and $m$, this PDF is shown in Figure \ref{fig:pdf_psi_Z}. Figures \ref{fig:pdf_psi_Y} and \ref{fig:pdf_psi_Z} look to be very similar, and we have confirmed this for other values of $\ell_{\mathrm{max}}$. That is, for a given $\ell_{\mathrm{max}}$, the distribution of the A-D statistic is the same for both the $Y$ and the $Z$ statistics. The distributions \textit{are} different for different values of $\ell_{\mathrm{max}}$.

\section{Results}

Having discussed the method for testing for the null hypothesis in the previous sections, in this section, we demonstrate that the method actually works. To do this, we break one of the assumptions in the null hypothesis. The easiest condition to break (in the sense that the new probability distribution is easiest to describe) is that of zero-mean. Previous studies \cite{cristian_mean} have looked at relaxing this condition, though they concentrate on somewhat larger values of $\ell$. They found that, at least in the range of multipoles they considered, the data seemed to be consistent with the zero-mean hypothesis. Here, we choose to break the condition of independence and normal distribution of the $a_{\ell m}$'s, mostly because that is usually posited as the reason behind the anomalies. But, we should emphasise that a similar analysis can be performed (in fact, more easily) with a non-zero mean.

Now, there is an infinite number of ways of breaking the independent, normally distributed hypothesis \cite{Abramo:2010gk}. We break it by deliberately masking the fiducial CMB sky about the equator. This masking breaks statistical isotropy and thus leads to a correlation between modes. The resulting probability distribution of the $a_{\ell m}$'s is difficult to analytically estimate, but it is clear that a greater degree of masking leads to a ``bigger" departure from the null hypothesis. Then, the strategy behind the demonstration is this:

\begin{enumerate}[topsep=0pt,itemsep=-1ex,partopsep=1ex,parsep=1ex]
\item Generate a set of fiducial CMB sky maps from a known set of $C_{\ell}$'s.
\item Generate $Y$ and $Z$ statistics using the $a_{\ell m}$'s of these maps.
\item Mask these maps to varying degrees and determine the resulting $a_{\ell m}$'s, and $Y$ and $Z$ statistics.
\end{enumerate}

The method can then be said to work if increasing the masking leads to a bigger departure from the null hypothesis (in the sense of the Anderson-Darling test applied to the $Y$ and $Z$ statistics).  Also, for zero masking, the distribution one gets with the CMB maps must correspond to $\Psi_Y(A^2, N, m)$ and $\Psi_Z(A^2, N, m)$ respectively.

As mentioned earlier, one of the things that we need to pick is the range of $\ell$'s that we will be considering. Because we are concentrating on low-$\ell$ anomalies, we start with the lowest relevant $\ell$ (\(\ell =2\)) and go up to an $\ell_{\mathrm{max}}$. For the rest of this section, let us choose \(\ell_{\mathrm{max}}=25\). From (\ref{eqn:lmax}), this corresponds to \(N=672\).

\begin{figure}[t]
    \centering
    \subfloat[No galactic mask]{{\includegraphics[width=0.48\linewidth]{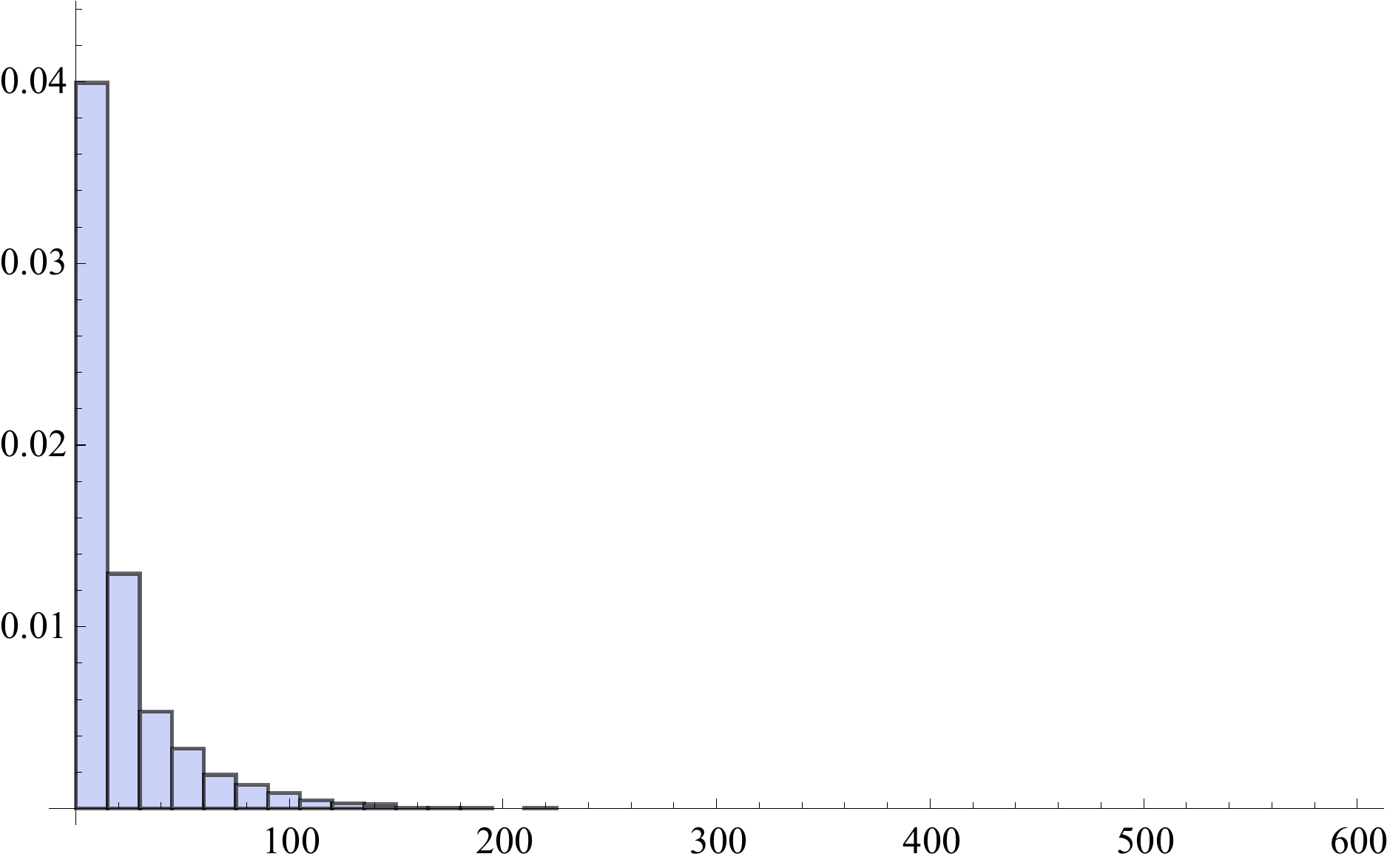}}}
    \quad
    \subfloat[5\% of pixels about the equator masked] {{\includegraphics[width=0.48\linewidth]{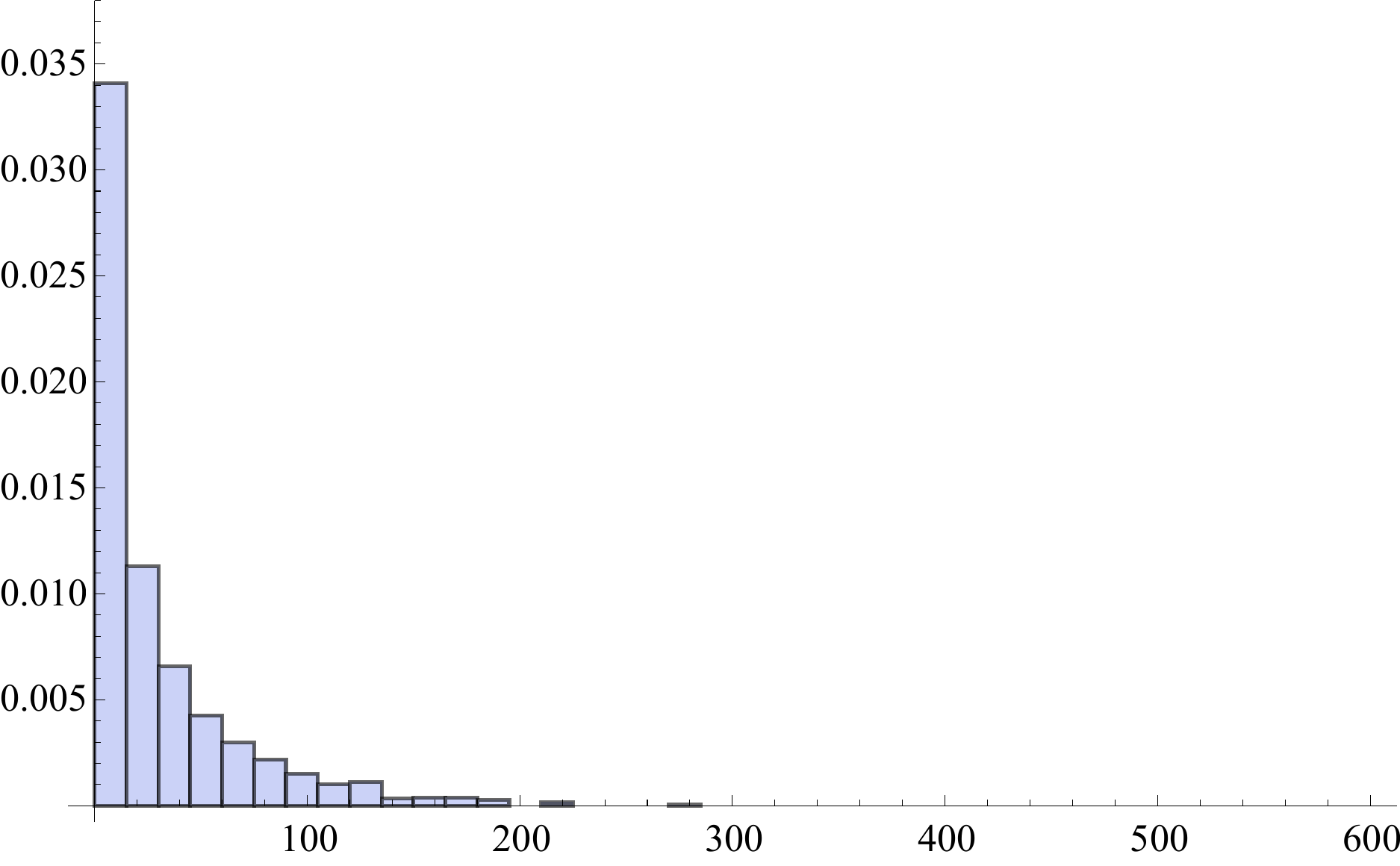}}}
	\newline
	\newline
    \subfloat[10\% of pixels about the equator masked] {{\includegraphics[width=0.48\linewidth]{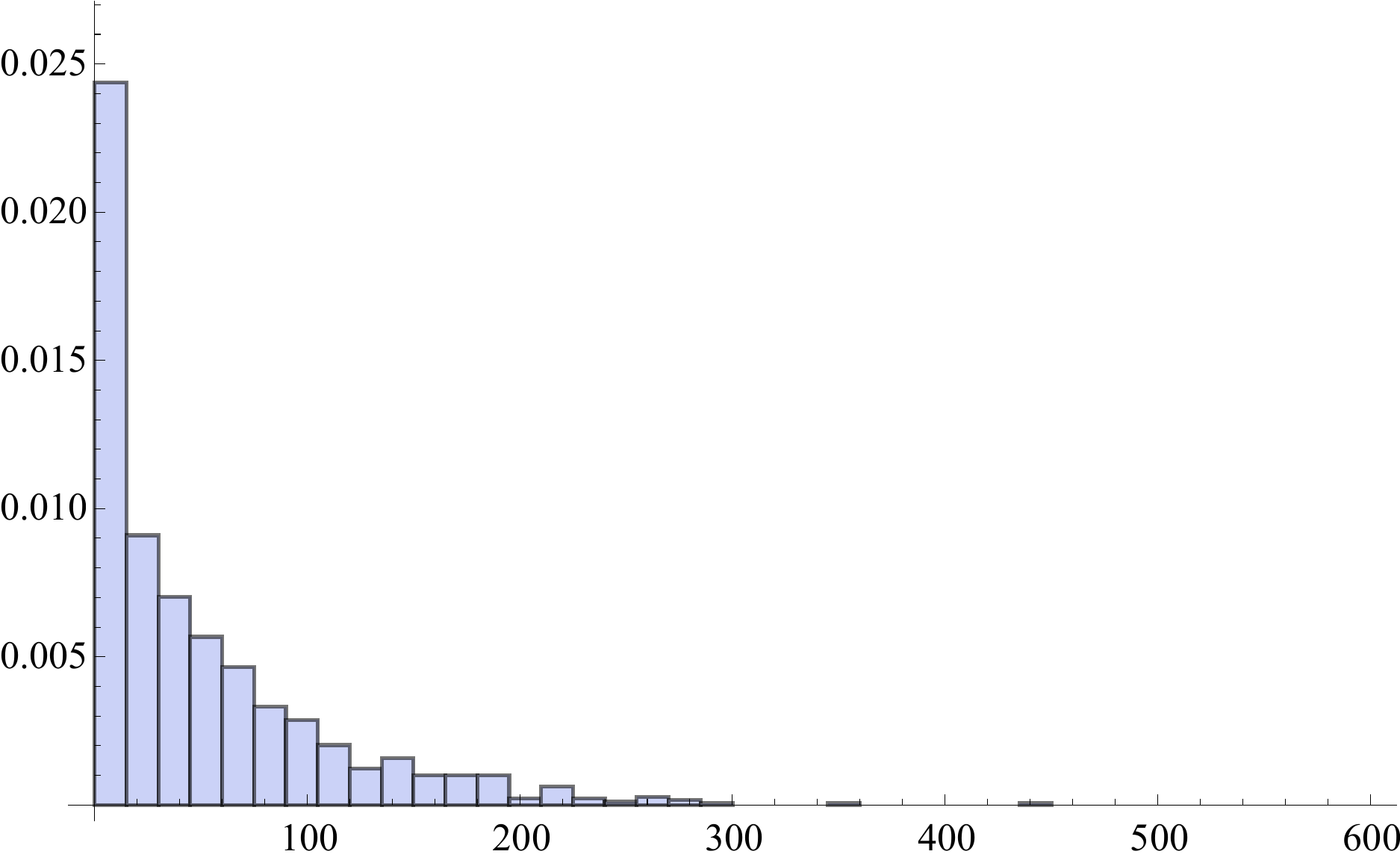}}}
    \subfloat[15\% of pixels about the equator masked] {{\includegraphics[width=0.48\linewidth]{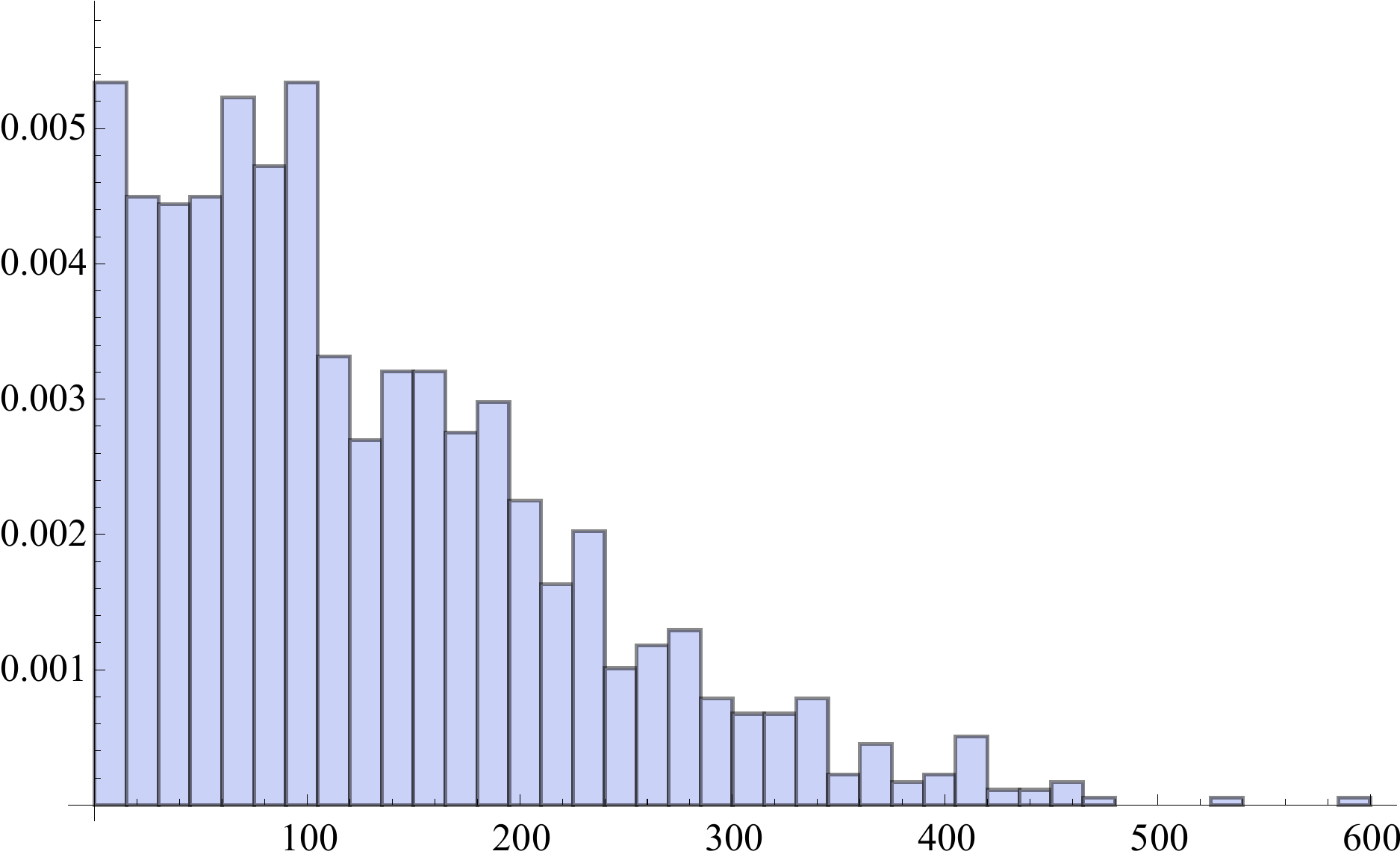}}}
    \caption{PDF for $\Psi_Y(A^2, N, m)$ for \(N=672, \, m=10^5\), and for different masks.}
    \label{fig:pdf_data_672}
\end{figure}

Therefore, what we now need to do is to generate $m$ realizations of $Y$ and $Z$ for each of the CMB maps described in the strategy above and compare this distribution with $\Psi_Y(A^2, N, m)$ for $Y$ and $\Psi_Z(A^2, N, m)$ for Z. We employ routines in HEALPix\footnote{\url{http://healpix.sourceforge.net/}} \cite{Gorski:2004by} to generate CMB maps from a given set of $C_{\ell}$'s, mask the maps, and then determine the corresponding $a_{\ell m}$'s. For the $C_{\ell}$'s, we use the Planck best-fit values, though, because this is for testing, any reasonable set would be sufficient. We consider four sets of maps: unmasked, and a mask of 5\%, 10\% and 15\% of the pixels about the galactic equator. We choose \(m = 10^5\), so that we can compare the distribution of the realized vector statistic with that in Figures \ref{fig:pdf_psi_Y} and \ref{fig:pdf_psi_Z}. We use C{}\verb!++! to generate the $Y$ and $Z$ statistics and {\large \textsc{mathematica}} to calculate the A-D statistic.

\begin{figure}[t]
    \centering
    \subfloat[No galactic mask]{{\includegraphics[width=0.48\linewidth]{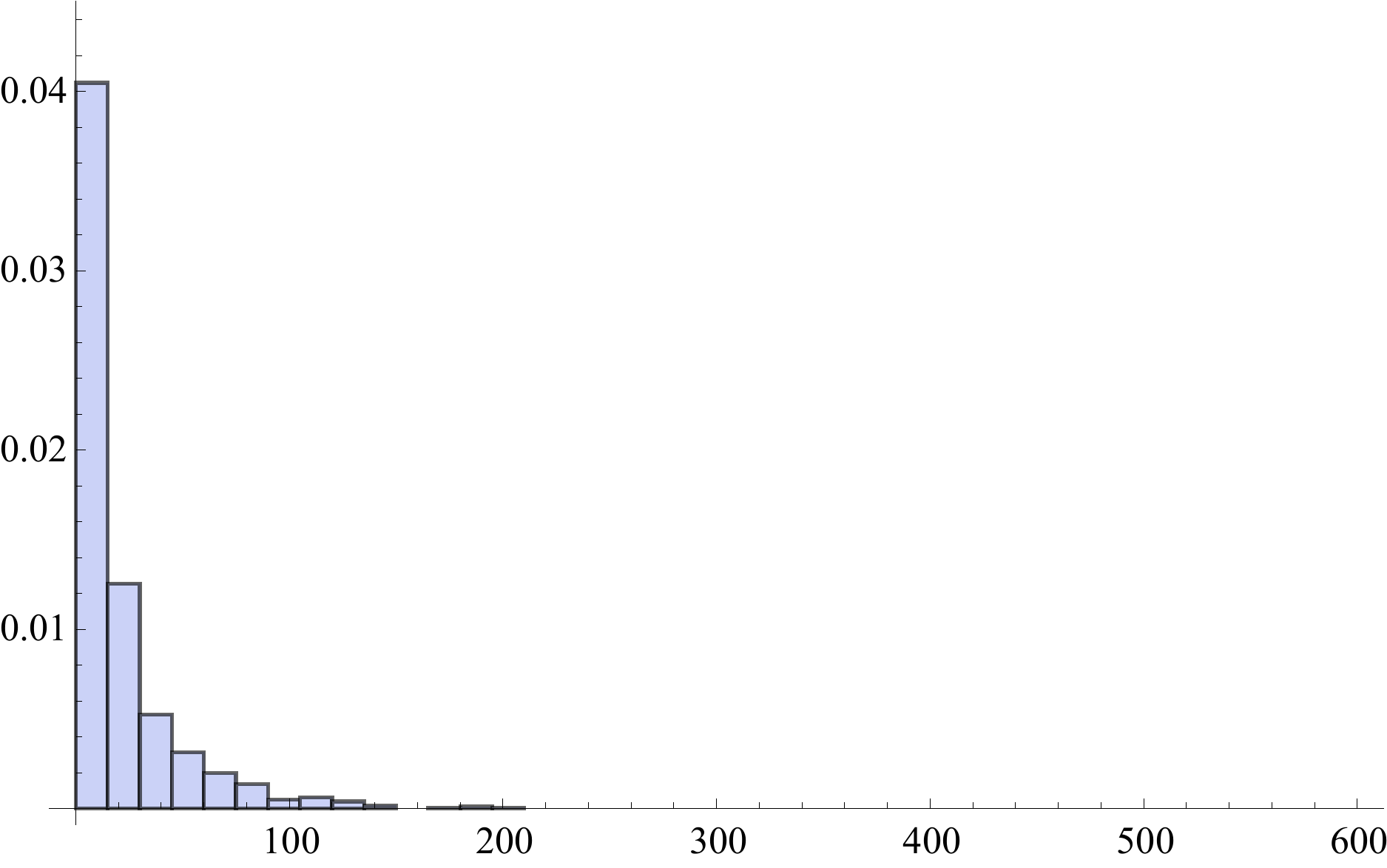}}}
    \quad
    \subfloat[5\% of pixels about the equator masked] {{\includegraphics[width=0.48\linewidth]{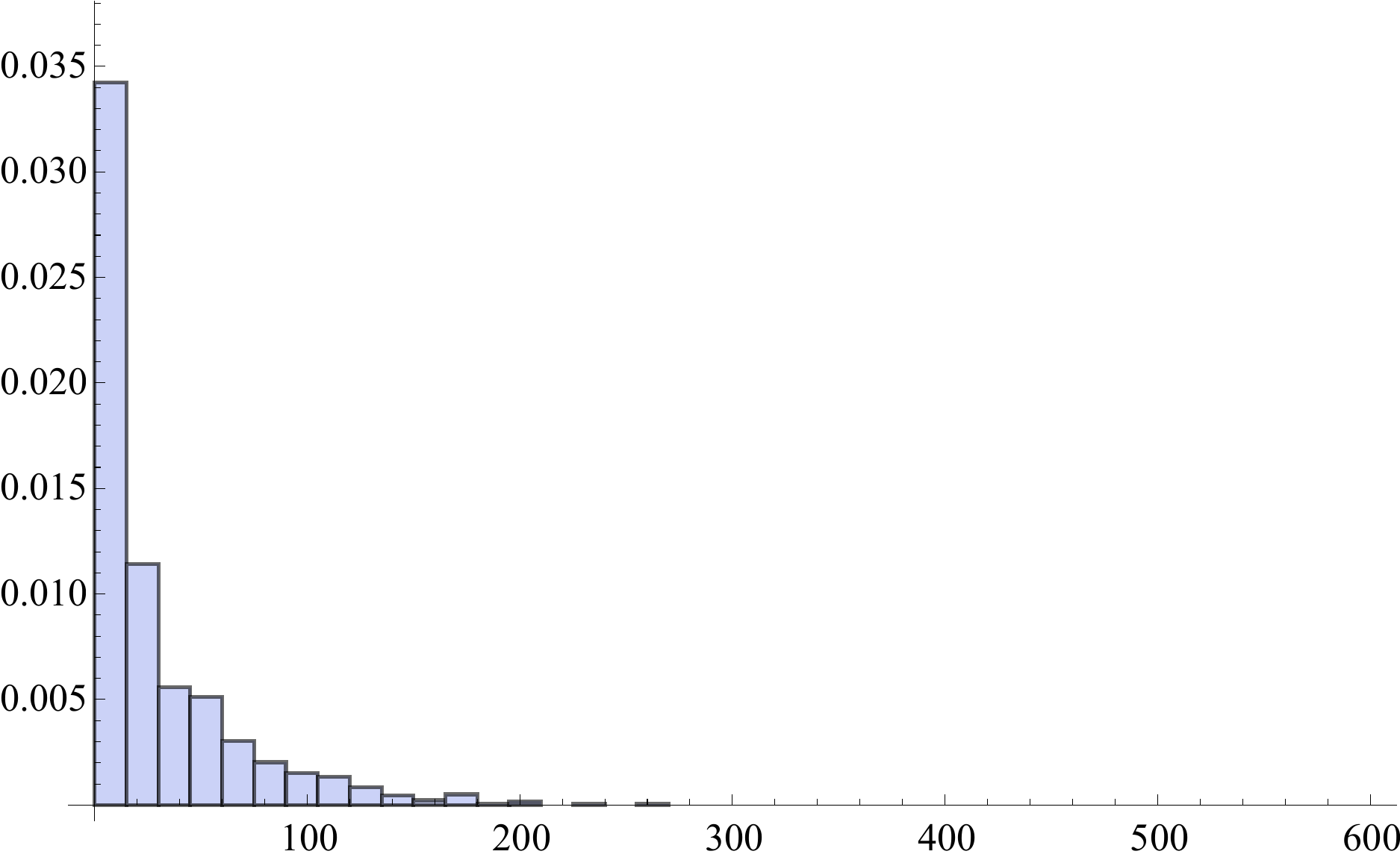}}}
	\newline
	\newline
    \subfloat[10\% of pixels about the equator masked] {{\includegraphics[width=0.48\linewidth]{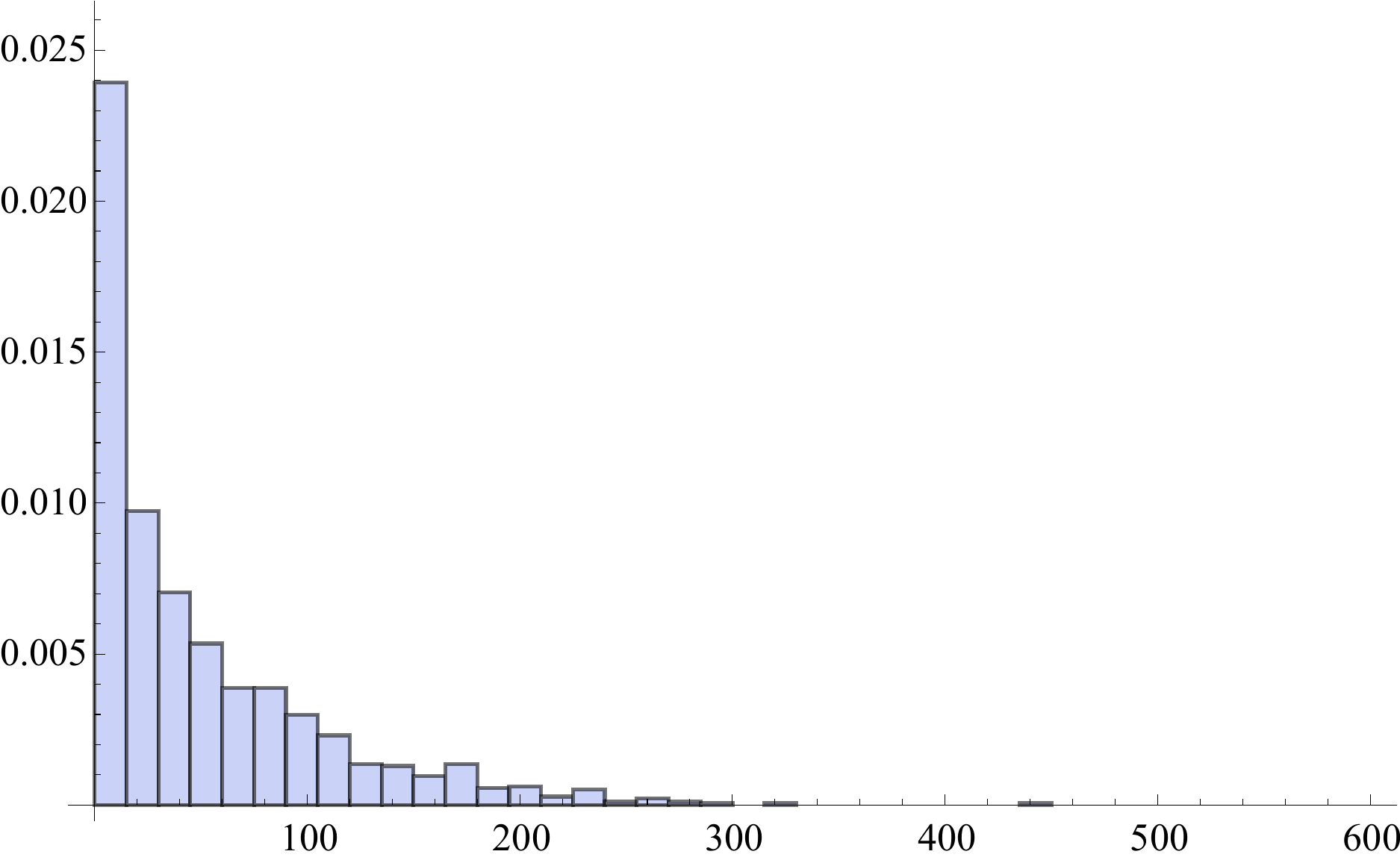}}}
    \subfloat[15\% of pixels about the equator masked] {{\includegraphics[width=0.48\linewidth]{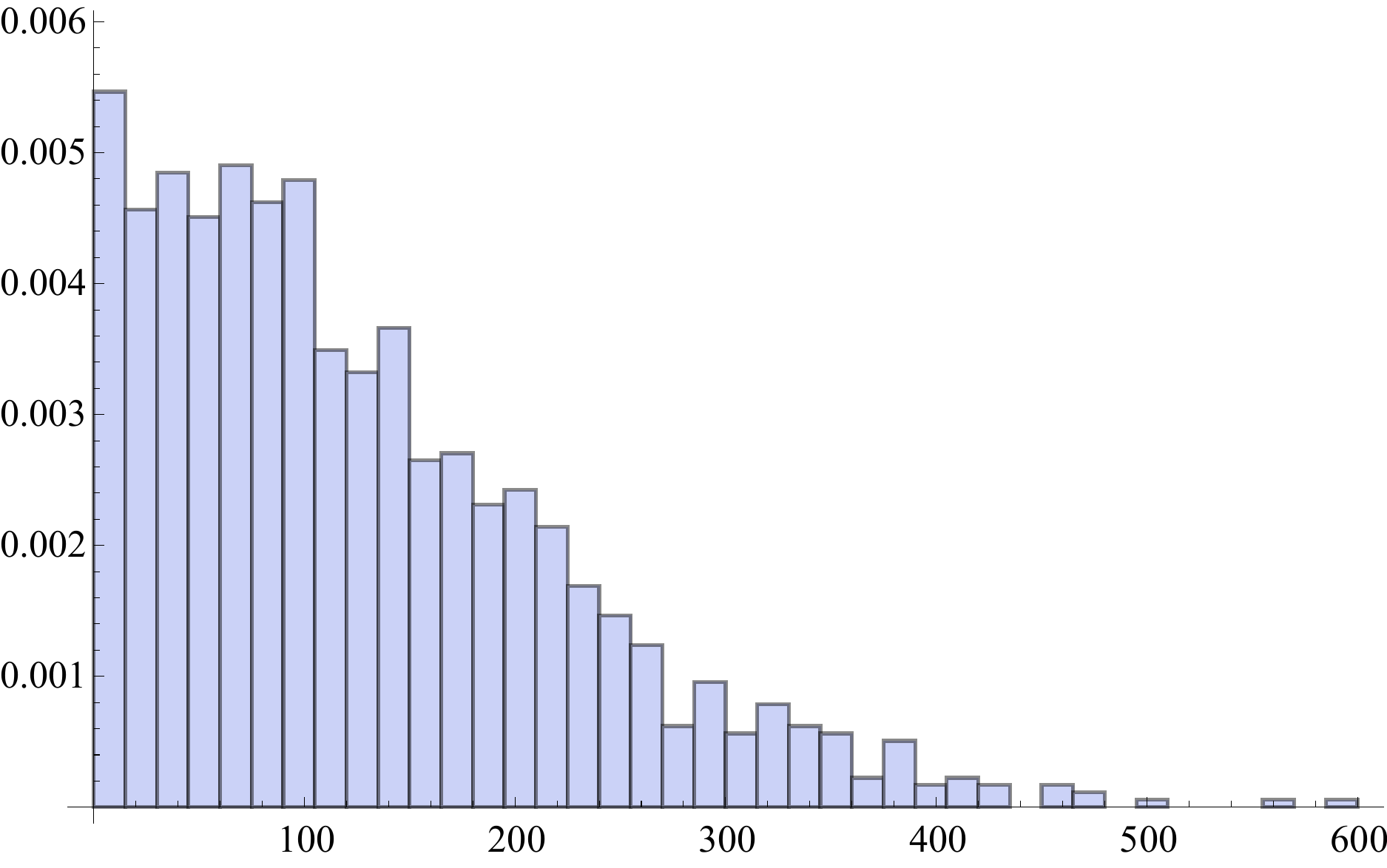}}}
    \caption{PDF for $\Psi_Z(A^2, N, m)$ for \(N=672, \, m=10^5\) and for different masks.}
    \label{fig:pdf_data_336}
\end{figure}

For the $Y$ statistic, the results are plotted in Figure \ref{fig:pdf_data_672}. As expected, the distribution for the unmasked sky [Figure \ref{fig:pdf_data_672} (a)] resembles that in Figure \ref{fig:pdf_psi_Y} to a very high degree, and the other three to a much lesser degree. Clearly, a bigger mask, and thus a bigger departure from statistical isotropy (and the null hypothesis), leads to a bigger departure of the distribution from that in Figure \ref{fig:pdf_psi_Y}. Similar results hold for the $Z$ statistic, plotted in Figure \ref{fig:pdf_data_336}. These plots are to be compared with those in Figure \ref{fig:pdf_psi_Z}. 

\section{Conclusion}

The last couple of decades have seen a tremendous amount of progress in the understanding of the large-scale structure of our Universe. Some parameters have been determined to several decimal places and some models have been ruled out to extremely high significance. Observationally, the only real challenge to this $\Lambda$CDM paradigm seems to be the large-scale CMB anomalies, of which many have been reported. The most important criticism levelled against the anomalies has to do with the fact that the anomalies are an \textit{a posteriori} phenomenon\textemdash one tests for anomalies after having ``looked" at the data. This is a fair criticism and in this paper we have proposed a method that addresses this very criticism. In a very general manner, we seek to test the null hypothesis that the $a_{\ell m}$'s are independent, zero-mean, normally distributed variables with an $m$-independent variance.

We consider linear ($Y$) and quadratic ($Z$) combinations of the $a_{\ell m}$'s, with randomized co-efficients. The probability distribution of these co-efficients is of a very specific form, but, depends only on the $C_{\ell}$'s. This choice greatly simplifies the PDFs of $Y$ and $Z$. Given a CMB map, the $Y$ and $Z$ distribution corresponding to the $a_{\ell m}$'s of the map can be determined. This distribution can be compared with the fiducial distribution for $Y$ and $Z$ (given in (\ref{eqn:dist_y_final}) and (\ref{eqn:dist_z}) respectively) and a high degree of incompatibility between the distributions would mean that the data are not well described by the null hypothesis.

To make this comparison between distributions, we have suggested a very slight modification of the Anderson-Darling test. Of course, other tests could also be used for this purpose. In order to demonstrate the usefulness of the test, we generated CMB maps with varying degrees of masking in them. This masking breaks statistical isotropy and thus results in a departure from the null hypothesis. We demonstrated that, firstly, for zero masking, the distribution of the Anderson-Darling test statistic is what we expect it to be. Secondly, increasing the masking did lead to distributions of the Anderson-Darling test statistic that were further and further removed from the distribution that arises out of the null hypothesis.

A few points to note regarding this method are: (i) Like most other ``goodness-of-fits" tests without an alternative hypothesis, this is a frequentist analysis. In particular, because of its very general and stochastic nature, the test may be susceptible to Type II Errors; that is, a failure to reject the null hypothesis. If we do have an alternative hypothesis, we can then compute the power of the test and make a quantitative statement about the probability of Type II errors. Or, indeed, do a Bayesian analysis. In the absence of this alternative hypothesis, a $p$-value compatible with the null hypothesis should \textit{not} be taken to mean that the data indicate that the null hypothesis is true.  (ii) In our analysis, we have assumed that the $C_{\ell}$'s are fixed numbers, but, at least from a Bayesian perspective, they themselves are random variables, with an associated variance. We don't see a way around this, because taking into account the stochastic nature of the $C_{\ell}$'s would make the analysis extremely complicated. Also, recall that the variance of $C_{\ell}$ is proportional to $C_{\ell}$ itself. Thus, for multipoles where the random nature of $C_{\ell}$ is most pronounced (that is, the lowest of the $\ell$'s), the value of $C_{\ell}$ is large to begin with. This partially alleviates the problem associated with assuming that the $C_{\ell}$'s are fixed numbers. (iii) Though we have concentrated on using the method to make statements about the $a_{\ell m}$'s, it is clear that our method works in general for any set of random variables that are hypothesized to be described by $H_0$. So, our method could be used to test $H_0$ in a variety of situations, becoming particularly useful when there are only a few realizations of several independent, \textit{non}-identically distributed Gaussian variables.

In a future publication, we hope to use our method and actual CMB data to quote $p$-values for the departure of the data from the null hypothesis. Planck is soon expected to release CMB polarization data, which can easily be incorporated into our analysis and should tell us more about the largest scales of the observable universe.

\section*{Acknowledgements}

It is a pleasure to thank Cristian Armendariz-Picon for his guidance throughout the project, and for a thorough reading of the draft. I would also like to thank Yabebal Fantaye for some crucial inputs, and Benjamin Wandelt and Eichiiro Komatsu for valuable feedback. Many thanks also to Stephan Lavavej for clarifications regarding pseudo-random number generation.

\bibliography{draftv4}
\bibliographystyle{unsrt}

\end{document}